# Wafer-scale fabrication of 2D van der Waals heterojunctions for efficient and broadband photodetection

*Jian Yuan[†,∥] Tian Sun,[†,∥] Zhixin Hu[‡,∥] Wenzhi Yu,[†] Weiliang Ma,[†] Kai Zhang[⊥], Baoquan Sun,[†] Shu Ping Lau,[#] Qiaoliang Bao,[§] Shenghuang Lin[#,*] and Shaojuan Li [†,*]*

[†]Institute of Functional Nano and Soft Materials (FUNSOM), Jiangsu Key Laboratory for Carbon-Based Functional Materials and Devices, Collaborative Innovation Center of Suzhou Nano Science and Technology, and Joint International Research Laboratory of Carbon-Based Functional Materials and Devices, Soochow University, Suzhou 215123, People's Republic of China

[‡]Center for Joint Quantum Studies and Department of Physics, Tianjin University, Tianjin 300350, People's Republic of China

[⊥]i-Lab, Suzhou Institute of Nano-Tech and Nano-Bionics, Chinese Academy of Sciences, Suzhou 215123, Jiangsu, People's Republic of China

[#]Department of Applied Physics, The Hong Kong Polytechnic University, Hung Hom, Hong Kong SAR, People's Republic of China

[§]Department of Materials Science and Engineering, and ARC Centre of Excellence in Future Low-Energy Electronics Technologies (FLEET) Monash University, Clayton, Victoria 3800, Australia.

[∥]These authors contributed equally to this work.

*Address correspondence to (S. Lin) shenghuanglinchina@gmail.com and (S. Li) sjli@suda.edu.cn

**ABSTRACT**

A variety of fabrication methods for van der Waals heterostructures have been demonstrated; however, their wafer-scale deposition remains a challenge. Here we report few-layer van der Waals $PtS_2/PtSe_2$ heterojunctions fabricated on a 2" $SiO_2/Si$ substrate that is only limited by the size of work chamber of the growth equipment, offering throughputs necessary for practical applications. Theoretical simulation was conducted to gain basic understanding of the electronic properties of the $PtS_2/PtSe_2$ heterojunctions. Zero-bias photoresponse in the heterojunctions is observed under laser illumination of 405 to 2200 nm wavelengths. The $PtS_2/PtSe_2$ heterojunctions show excellent characteristics in terms of broadband photoresponse and high quantum efficiency at infrared wavelengths, as well as a fast response speed at the millisecond level. The wafer-scale production of 2D heterojunctions in this work accelerates the possibility of 2D materials for applications in the next-generation energy-efficient electronics.

Heterojunctions, firstly invented in 1963 by Herbert Kroemer[1], have been essential building blocks for electronic and optoelectronic devices in the current semiconductor industry. Uses of such structures have been envisioned in nearly every type of semiconductor devices, including biomedical devices, bipolar transistor, photodiodes, light-emitting diodes, and solar cells. As one of the most important existing optoelectronic devices, the photodiodes obviously play an important role in the applications of video imaging, optical communication, remote control and night vision[2, 3]. Recent studies have shown that heterostructures based on two-dimensional (2D) materials, which contains atomically sharp interfaces, can indeed be competitive with or even superior to conventional bulk semiconductor based junctions[4, 5, 6]. Encouragingly, 2D transition metal dichalcogenides (TMDs), one important member from the family of atomically thin van der Waals materials, has been widely studied and proved to be of great potential for the applications of future optoelectronics owing to their outstanding electronic, optical, mechanical properties and the strong light–matter interactions[7-9]. TMDs have extended bandgap tunability through composition[10], thickness[11, 12] and possibly even strain control[13] offering infinite flexibility to design 2D junctions[4, 14-18], which has been used in photovoltaics, photodiodes and light emitters[11, 15, 19], and could overcome some of the existing problems in conventional junction devices[20].

Therefore, driven by the diversity and considerable wide coverage properties of TMDs materials, artificial 2D van der Waals junctions have been fabricated using either homogeneous or heterogeneous 2D materials [17, 21]. The formation of 2D homojunction can be formed by chemical/gate-induced electrical doping in the same nanoflake which lacks stability or makes device structures complicated[19, 22, 23]. The fabrication of 2D heterojunctions, however, mainly extends directly to materials produced by exfoliation from the bulk counterpart using a variety of techniques[6, 17], but indirectly to those that can be deposited onto targeted substrates with facile control for the practical realization of high-volume manufacturing. Controllable large-scale integration of more than one type of 2D materials onto a single substrate still remains a challenge. Previously, we have demonstrated the scalable production of periodic patterns of few-layer $WS_2/MoS_2$ heterojunction arrays by a thermal reduction sulfurization process[24], however, the TMDs/TMDs heterostructures reported display a spectral response that is mainly limited to visible wavelengths by the intrinsic bandgap of the constituting materials[25].

Recently, the group 10 metal based TMDs have attracted intense interests for the widely tunable bandgap, large electrical conductivity and high air stability, which can well make up the drawbacks of graphene (zero bandgap), other TMDs semiconductors (relatively large bandgap), and black phosphorus (poor air stability). Representative examples of this family are platinum diselenide ($PtSe_2$) and platinum disulfide ($PtS_2$) that can be synthesized via a single step, *i.e.* direct selenization (or sulfuration) of the Pt substrate[25, 26]. Layered $PtSe_2$ has a tunable bandgap ranging from 1.2 eV (monolayer) to semimetal (bulk)[25], while layered $PtS_2$ has a bandgap that varies from 1.6 eV in monolayer to 0.25 eV in bulk[27]. The large spread of bandgaps from visible to mid-infrared is similar to that of layered black phosphorus (BP)[28], revealing potential for infrared electronic application. Compared to other extensively studied TMDs, such as $MoX_2$, $WX_2$ (X is S, Se, or Te) *etc*, field-effect transistors based on $PtSe_2$ layers exhibit higher mobility at room temperature, suggesting that this material is applicable for high-speed electron transport devices[29]. Furthermore, few-layer $PtSe_2$ nanosheets have been used to make infrared photodetectors with good photoresponsivity and fast response[29-32]. Most recently, bilayer $PtSe_2$ field effect transistors were demonstrated for broadband mid-infrared photodetection with a high photoresponsivity of ~4.5 AW$^{-1}$ and response speed at the millisecond level[33]. Previous research on few-layer $PtS_2$ phototransistors also demonstrates high photoresponsivity and photoconductive gain due to the existence of trap states[34]. Although the optoelectronic properties of individual 2D group-10 TMDs have been preliminary studied, the comprehensive study of van der Waals like interlayer coupling of two different group-10 TMDs and their 2D heterojunctions, which can dramatically affect the band-structures and optoelectronic properties of 2D TMDs, are still lacking. By combining their advantages of ultrahigh stability and infrared light absorption, group-10 TMDs based 2D heterojunctions could be promising for low-power and high speed optoelectronic devices at infrared wavelengths.

Herein, we report a simple strategy to create few-layer $PtS_2/PtSe_2$ heterojunctions on the surfaces of planar substrates to enable wafer-scale manufacturing of 2D heterostructure arrays based on group-10 TMDs. Such 2D heterojunctions can be formed directly through selenization (or sulfuration) of the Pt substrate by changing the Se source from S on the Pt layer using an ambient pressure conversion process. The coverage, size, and shape of $PtS_2/PtSe_2$ film can be controlled as desired, and their thicknesses were determined by the thicknesses of the pre-deposited Pt film. Methods based on above process are naturally compatible with modern planar technologies, and they offer throughputs necessary for practical applications. Theoretical simulations were conducted using the Vienna ab initio simulation package to gain basic understanding of the electronic properties of the $PtS_2/PtSe_2$ heterojunctions. Zero-bias photoresponse in the heterojunctions is observed under laser illuminations of 405 to 2200 nm wavelengths. Upon optical illumination, the $PtS_2/PtSe_2$ heterostructures show high photoresponsivity and quantum efficiency at infrared wavelengths with lower bounds for the external quantum efficiencies (EQE) being 1.2% at 1064 nm, 0.2% at 1550 nm and 0.05% at 2200 nm , as well as fast response speed at the millisecond level. These heterojunctions show great promise for broad-band photodetection.

## RESULTS AND DISSCUSSION

The vertical heterojunction arrays (see Figure 1a and 1b for microscope images) were obtained on an oxide silicon wafer with 300 nm thick silicon dioxide. They consist of few-layer $PtS_2$ and $PtSe_2$ films that were synthesized via direct sulfuration or selenization of the Pt substrate by changing the Se source from S on the Pt layer using an ambient pressure conversion process (see Methods for more details). Our fabrication process enables manufacturability of wafer-scale production of 2D heterojunctions for optoelectronic applications, and the maximum sample size obtained in our experiment is up to 2" in diameter limited by the work chamber of the equipment. From the high-resolution optical microscopy images (Figure 1b), we can observe that the $PtS_2$ and $PtSe_2$ sheets are well-defined with very uniform contrast, irrespective of single $PtS_2$ or $PtSe_2$ sheet or their overlap region, which is attributed to the delicate control of growth process, proper thermal budget and homogeneous thickness of $PtS_2$ and $PtSe_2$ sheets. The size of $PtS_2$ sheets and $PtSe_2$ sheets were designed to be $100 \times 120 \ \mu m^2$ and $100 \times 100 \ \mu m^2$, respectively. Atomic force microscopy (AFM) was used to probe the detailed surface morphology and the thicknesses of $PtS_2$, $PtSe_2$ sheets, and their vertical heterojunctions (Figure 1c). The AFM height profile indicates that the thickness of $PtS_2/PtSe_2$ heterojunction is ~5.1 nm.

We have also measured the pure $PtS_2$ and $PtSe_2$ regions, respectively. The thicknesses of $PtS_2$ and $PtSe_2$ are measured to be ~2.5 and ~2.7 nm, corresponding to five layers of $PtS_2$ and five layers of $PtSe_2$, respectively[25].

Raman spectroscopy was utilized to identify and characterize the obtained $PtS_2$/$PtSe_2$ heterojunction films. Figure 1d shows the polarization-dependent Raman spectra of $PtS_2$/$PtSe_2$ heterojunction excited by a 633 nm laser. Both the feature peaks of $PtS_2$ and $PtSe_2$ are observed in the spectra. The two feature peaks at ~176 $cm^{-1}$ and ~210 $cm^{-1}$ correspond to the $E_g$ and $A_{1g}$ Raman active modes of $PtSe_2$, respectively. The $E_g$ mode is an in-plane vibrational mode of Se atoms moving away from each other within the layer, while the $A_{1g}$ mode is an out-of-plane vibration of Se atoms in opposing directions. There is also a small peak located at ~230 $cm^{-1}$ which can be attributed to a longitudinal oscillation (LO) mode, similar to those observed in $HfS_2$, $ZrS_2$ and $CdI_2$ [32, 33]. For $PtS_2$, three main peaks at about 307, 335, and 340 $cm^{-1}$ can be assigned to $E_g^1$, $A_{1g}^1$, and $A_{1g}^2$ phonon modes, respectively. On the basis of previously theoretical prediction and experimental observations[35], we know that these two out-of-plane modes, $A_{1g}^1$ and $A_{1g}^2$ are observable in the polarization-dependent Raman spectra of $PtS_2$, while their intensities are polarization-dependent. This phenomenon can also be found in our experiment as shown in Figure 1d, the difference is that the intensity disparity between these two modes are not as obvious as the previously reported results because the samples in our experiments are formed with fine grains, as verified by X-ray diffraction (XRD), transmission electron microscopy (TEM) and Scanning TEM (STEM), which will be elucidated in the following part. Furthermore, Raman mapping images (inset of Figure 1d) indicate that the overlap part of the $PtSe_2$ and $PtS_2$ sheet are well-separated, where the $PtSe_2$ sheet stacks on the top of the $PtS_2$ sheet. The crystal structures of $PtS_2$ and $PtSe_2$ films were further explored by XRD, as shown in Figure S1a. The layered $PtS_2$ and $PtSe_2$ film can be viewed as cleaved from the (0001) surface of the bulk $PtS_2$ and $PtSe_2$, where one Pt atom layer is sandwiched between two S or Se layers ($PtS_2$: JCPDS PDF No. 01-070-1140, $PtSe_2$: JCPDS PDF No. 01-088-2280). The main XRD diffraction peaks of $PtSe_2$ films are observed at about 17.6°, 34.6°, and 54.6°, which can be indexed to the (001), (002) and (003) crystal planes of $PtSe_2$, respectively, suggesting that $PtSe_2$ film grows along c-direction with (001) as bottom plane, which agrees very well with the previous report[35,36]. Similar results were found for the $PtS_2$ film. It is noteworthy that the diffraction peaks of Pt (111) disappeared after selenization or sulfuration which indicates a complete conversion of Pt into $PtSe_2$ and $PtS_2$.

The X-ray photoelectron spectroscopy (XPS) was also used to determinate the elemental binding energies of $PtS_2$, $PtSe_2$ sheets, and their vertical heterojunctions, as shown in Figure 1e and Figure S1(b). XPS spectra of the Pt 4f, Se 3d and S 2P were acquired on the heterojunction region. The measurement results for the binding energies clearly demonstrate the formation of $PtS_2$ and $PtSe_2$. The peak positions at 54.39 and 55.19 eV correspond to the binding energy of $Se^{2-}$. Besides, two other peaks located at 73.9 and 77.3 eV of Pt 4f spectrum can be assigned to the Pt $4f_{7/2}$ and Pt $4f_{5/2}$. The peaks at 162.3 and 163.6 eV of S 2p spectrum are ascribed to binding energy of Pt-S bonds between adjacent Pt atoms and S-S bonds, respectively, similar to the previous reports[36, 37]. To check the purity of the materials, the corresponding Pt, S and Se element mapping images were measured by energy dispersive X-ray spectroscopy, as displayed in Figure 1f, which confirms that all the elements are homogeneously distributed throughout the entire structure.

To further assess the microstructure, crystallinity, and elemental composition of the as-grown $PtS_2$/$PtSe_2$ vertical heterojunctions, the samples were transferred onto copper grid using the poly(methyl methacrylate) (PMMA) assisted transfer method[38] and investigated by TEM and STEM, as shown in Figure S2. The low-magnification TEM image (Figure S2b) indicates that the $PtSe_2$ and $PtS_2$ nanosheets have good uniformity and continuity across the whole platelet. The inset in Fig. S2c shows the selected area electron diffraction (SAED) pattern of $PtSe_2$ nanosheet, which confirms the obtained sample is polycrystalline and the four distinguished red dashed circles are assigned to (001), (101), (111) and (201) planes with lattice spacings of 5.15, 2.77, 1.55 and 1.75 Å, respectively. The inset in Fig. S2d shows the SAED pattern of $PtS_2$ nanosheet, which confirms the obtained sample is polycrystalline and the four distinguished red dashed circles are assigned to (101), (102), (111) and (202) planes with lattice spacings of 2.61, 1.93, 1.65 and 1.29 Å, respectively. The high-resolution TEM (HRTEM) image of the $PtSe_2$ nanosheet in Figure S2e reveal clear lattice fringes with a lattice spacing of 0.287 nm corresponding to (101) facets of $PtSe_2$ nanosheet, while for the $PtS_2$ nanosheet, the lattice space is 0.261 nm as shown in Figure S2f.

To gain basic understanding of the electronic properties of the $PtS_2/PtSe_2$ heterojunctions, theoretical simulations were conducted using the Vienna ab initio simulation package. Details of the simulation process is shown in Methods. The atomic geometries for layered $PtSe_2$, $PtS_2$ and coupled structure are shown in Figure 2 a-c. The 1T phase is selected since it is the most stable structure. The thickness of 5L-$PtSe_2$ and 5L-$PtS_2$ are 27.6 Å and 24.4 Å, comparable to the experimental value of 2.7 nm for $PtSe_2$ and 2.5 nm for $PtS_2$, respectively. Distances between adjacent Pt atoms in $PtSe_2$ and $PtS_2$ are 3.74 Å and 3.58 Å, respectively. When modeling coupled structure of $PtSe_2$ and $PtS_2$, the lateral lattice parameter of $PtS_2$ is slightly expanded in order to fix in the unit cell. Both 5L $PtSe_2$ and $PtS_2$ in 1T phase are observed to be semiconductors in experiments[29, 34, 36, 37]. Figure 2 d-f show the simulated band structures. 5L-$PtSe_2$ has an indirect bandgap of 0.21 eV. The conduction band minimum (CBM) is settled between Gamma and M point, while the valance band maximum (VBM) is slightly offset from Gamma point. 5L-$PtS_2$ is also a semiconductor with indirect bandgap of 0.89 eV. Calculated bandgaps are highly consistent with the experimental observation[27]. For the band structure of coupled system shown in panel f, the calculated band structure is only 0.03 eV, which seems to indicate a much more metallic feature, corresponding to a wide spread wavelength range. The plotted bands are a mixture of states from both $PtS_2$ and $PtSe_2$, which cannot represent the electronic properties of $PtS_2$ or $PtSe_2$ layers separately. In order to give a better estimation of the bandgap for each type of material, the projected density of states (PDOS) for Pt and Se/S atoms are also shown adjacent to band structures. By comparing the PDOS before and after the coupling, it can be found that the bandgap of $PtS_2$ is shrunk to half of its original size. Part of the shrink is caused by changing the lattice parameter. The PDOS of $PtSe_2$ is almost unchanged, indicating a limited response to the coupling. Both $PtSe_2$ and $PtS_2$ layers in the coupled system are still semiconductors since clear bandgaps are present in the PDOS plot. The VBM of coupled system is contributed by the $PtSe_2$ while the CBM is contributed by $PtS_2$ as shown in Figure S3 of Supporting Information. The shift of energy level respecting to the vacuum level is shown in Figure 3a. The conduction and valence band positions of 5L $PtS_2$ is at around –5 eV and –5.9 eV, respectively, while, the conduction and valence band positions of 5L $PtSe_2$ is located at -4.81 eV and -5.01 eV, respectively. The bandgap of $PtS_2$ decreases by 0.36 eV when its lattice parameter is stretched to the value of $PtSe_2$.

The coupled system has its conduction band and valance band slightly lifted up from their original levels, creating a tiny bandgap of 0.03 eV.

To further determine the band offset at the PtS$_2$/PtSe$_2$ interface, we performed ultraviolet photoelectron spectroscopy (UPS) measurements to test the band edge energies (Supporting Information, Figure S4). The results show that PtS$_2$ exhibits a higher work function (~ 4.99 eV) than that of PtSe$_2$ (~ 4.93 eV). The work function offset (60 meV) between PtS$_2$ and PtSe$_2$ should block electron flow from the PtS$_2$ film to the PtSe$_2$ film after contact. Note that the work functions derived from UPS mesearments results may deviate from actural values because the material is etched by Ar iron-beam before UPS characterization (see supporting information Figure S4 for details). Recent research results have shown that the bandgap of PtSe$_2$ can be efficiently changed through defect engineering via Ar plasma treatment[33]. The influence of plasma treatment on work function needs further investigation, which is out of the scope of this study. To this regard, kelvin probe force microscopy (KPFM) measurements were also performed to evaluate the energy offset at the PtS$_2$/PtSe$_2$ heterojunction interface and identify direction of the photocarriers transportation. The representative KPFM image under illumination of white light is shown in Figure S5a of Supporting Information. Figure S5b shows the corresponding surface potential taken along the dashed line in Figure S5a. The surface potential of the PtS$_2$ film is lower than that of the PtSe$_2$ film, which manifests PtS$_2$ has higher work function than PtSe$_2$. The work function difference between the two materials is approximately 35 meV (blue dashed line in Figure S5a). This disparity with UPS measurements result (60 meV) is because of the different environment in measurement ( *i.e.* in air for KPFM measurement and in vacuum for UPS measurement), and defect induced bandgap change via Ar plasma treatment (as elucidated above). Combined with our theoretical calculation results, the schematic band diagrams in 5L PtS$_2$/5L PtSe$_2$ heterojunction can be depicted, as shown in Figure 3b. After contact, the Fermi levels in these two materials are shifted so as to satisfy the equilibrium condition due to the work function mismatch [39, 40]. The electrons diffuse from PtSe$_2$ to PtS$_2$ and forms a built-in electric field, which leads to band bending between these two materials and facilitates separation of photo-excited carriers. Figure 3b also illustrates the transfer of photo-excited electron-hole pairs under light illumination. It is proposed that at least three processes can contribute to the photocurrent. The photogenerated electrons in PtSe$_2$ at the interface of heterojunction are swept into PtSe$_2$ by the built-in electric field, while holes are blocked by the high barrier and accumulate at the interface. PtS$_2$ can also absorb light and generate electron-hole pairs. Moreover, carriers in the conduction band of the PtS$_2$ may be injected into PtSe$_2$, which will contribute more electrons to increase the photocurrent.

Figure 3c, d show the schematic illustration of a photodetector based on the PtS$_2$/PtSe$_2$ heterostructure and the corresponding current-voltage (*I-V*) characteristics in dark and under illumination. The device shows typical junction characteristics as seen from Figure 3d. At positive bias voltage, the current gradually enhances with the applied voltage ($V_{SD}$) due to the increase of carrier density. Furthermore, we found that the device can operate at no applied bias under infrared light illumination (1064 nm), as seen from Figure 3d and Figure 3e. Figure 3e shows the typical photocurrent switching performance of the heterostructure at the infrared wavelength of 1064 nm ($V_{SD} = 0$ V, $V_G = 0$ V). Obvious photoresponse is observed under different laser powers. This kind of self-powered operation behavior can have large numbers of applications like powerless communications, biological and chemical imaging in a wireless healthcare platform, and so on. A rectification ratio in the range of 15~20 was found for dozens of devices we have measured, which is similar with the values in recently reported TMDs-based junctions[31, 41, 42, 43]. Furthermore, this relative weak rectifying behaviour indicates thermionic and/or tunneling electron currents can be present in our heterojuction.

Responsivity is an important figure of merit for a photodetector and reflects its sensitivity to the incident light. In order to normalize the incident power by the active area of the device, here we consider that the active area is the overlap region of the heterojuction, the responsivity (R = $I_{photo}/P_{incident}$, $I_{photo}$ is the photocurrent and $P_{incident}$ is the incident power on the device active area) can be obtained. Figure 3f shows the dependence of photocurrent and responsivity on incident light power at zero-volt state. The photocurrent increases linearly while promoting the light power, while the responsivity decreases as the increase of incident power, in consistant with the behaviors of other TMDs-based heterojuctions reported before[24,44].

With the aim to demonstrate the capability of the broadband ligh detection of our device, a series of photoelectric measurements were performed at different wavelengths.

The temperal photoresponse with various wavelengths under the same light power (15 mW) is shown in Figure 4a. The device can be effectively switched ON and OFF while the light source is turned on and off even in the wavelength of 2200 nm (See the power-dependent photoresponse at 2200 nm in Figure S6, Supporting information). Moreover, the spectral photoresponse of the junction (Figure 4b) generally follows the absorption spectrum of 5L PtSe$_2$. The wavelength-dependent responsivity (Figure 4c) under the same light power shows a maximum at visible light illumination of 532 nm (15 mW) and then decreases as the photon energy decreases. The external quantum efficiency (EQE) represents the ratio of the number of collected charge carriers to the number of incident photons. It can be calculated by the formula $\mathrm{EQE} = \mathrm{I}_{photo} / P_{incident}(\hbar c / e\lambda)$, where $\lambda$ is the incident wavelength, $e$ is the electronic charge，$\hbar$ is planck constant, $c$ is the speed of light[15, 17]. The EQE values as a function of wavelength are shown in Supporting Information Figure S7a. EQE is caculated to be around 7.1% at 1064 nm, 1.2% at 1550 nm and 0.2% at 2200 nm. Note that these EQE values may be overestimated since we considered only the overlap region of the heterojuction absorbs light. However, we found that if we consider the active area to be the entire semiconducting region, the calculated EQE values decreases by a factor of six, that is 1.2% at 1064 nm, 0.2% at 1550 nm and 0.05% at 2200 nm, which are the lower bounds for the EQE.

We further investigate the response speed of PtS$_2$/PtSe$_2$ heterojunction photodetector by shining the device with pulsed light that is modulated by an optical chopper as shown in Figure 4d. The response of the heterojunction is very fast with the rise and decay times to be 66 ms and 75 ms, respectively (Supporting Information, Figure S10a). The photoresponse of the PtS$_2$/PtSe$_2$ heterojunction remains highly stable after 60 days under the same illumination conditions, showing the excellent stability of the device in air at room temperatrue (Supporting Information, Figure S10b), which is very important for the practical implementation of this device in long-term operation. The performance of our PtS$_2$/PtSe$_2$ heterojunction devices is compared to other TMDs-based heterojunctions, as summarized in Table 1. The overall performance of our device is superior to existing MX$_2$/MX$_2$ (M = Mo, W; X = S, Se, Te) heterostructures in terms of working spectral range with comparable response time. We attribute the excellent photoresponse of PtS$_2$/PtSe$_2$ heterostructrue to the small bandgap and enhanced light absorption in the coupled system.

Note that the 2D material quality remains far from optimal as evidenced from the TEM measurements, the response time of the hybrid system is only comparable to pristine $PtS_2$ photodetector, which is mainly limited by the existence of trap states[34], and may be further improved by optimizing the material quality and growth process.

**CONCLUSION**

In summary, wafer-scale 2D heterojunctions based on few-layer $PtS_2$ and $PtSe_2$ film were produced on a 2" $SiO_2$/Si substrate. Theoretical calculation along with the KPFM, UPS experimental verification were conducted to evaluate the electronic properties of the $PtS_2$/$PtSe_2$ heterojunction. Zero-bias photoresponse in the heterojunctions is observed under laser illumination of different wavelengths. The self-driven $PtS_2$/$PtSe_2$ heterojunctions show excellent characteristics in terms of wide photoresponse range from 405 to 2200 nm, high EQE at infrared wavelengths and a fast response speed. The speed of the device may be further improved by optimizing the material quality and growth process. The results demonstrated in this work may open up more possibilities toward infrared optoelectronic applications based on 2D materials.

**METHODS**

**Synthesis of the $PtS_2$/$PtSe_2$ heterojunctions.**

A two-step chemical vapor deposition approach was used to produce wafer-scale $PtS_2$/$PtSe_2$ heterojunctions in a double heating area furnace, similar to our previous report[24]. In step one, photoresist was first spin-coated on the $SiO_2$/Si substrate and exposed to form the array of periodic square holes. Subsequently, 0.8 nm Pt was deposited by electron beam evaporation to the corresponding square holes. Then the $SiO_2$/Si substrate with Pt was placed into the quartz tube at the downstream and heated to 600 °C and sulfur powder was then put at the upstream with 130 °C. Argon (Ar ) was introduced as the carrier gas (flow rate: 60 sccm). The $SiO_2$/Si substrates with Pt was maintained at 600 °C for 2 hours to form $PtS_2$. In step two, photoresist was spin-coated on the $SiO_2$/Si substrate with $PtS_2$ array and exposed again to form the periodic rectangle hole array on part of the $PtS_2$ sheets. After that, 0.8 nm Pt was then deposited by electron beam evaporation to form the periodic rectangle Pt arrays. Following, similar to the process of producing $PtS_2$ mentioned above, the furnace temperature was kept at 450 ºC for 2 hours to prepare PtSe$_2$.

**Fabrication of the PtS$_2$/PtSe$_2$ heterojunctions devices.**

UV lithography was used to define the source-drain electrodes pattern onto the PtS$_2$/PtSe$_2$ heterojunctions and then 5 nm Ti/80 nm Au was deposited through electron beam evaporation technique to form the source-drain electrodes.

**Characterization of PtS$_2$/ PtSe$_2$ heterojunctions.**

The morphology and structure of the synthesized materials were characterized by optical microscopy (OM, Olympus DX51), scanning electron microscopy (SEM, (FEI Quanta 200 FEG, acceleration voltage: 5 30 kV)), atomic force microscopy (AFM, Digital Instrument Nanoscope IIIA), transmission electron microscopy (TEM, FEI Tecnai F30, acceleration voltage: 200 kV), and micro-Raman spectroscopy (Horiba, LabRAM HR-800). X-ray photoelectron spectroscopy (XPS) measurement was performed on a KRATOS AXIS Ultra DLD equipment (KRATOS Analytical C.O.). Photoelectric measurements were performed on a probe station (Cascade M150) equipped with a semiconductor property analyzer (Kiteley 4200) in ambient conditions. More photoelectrical properties of the devices are described in the Supporting Information.

**Simulations of the energy band of the PtS$_2$ and PtSe$_2$.**

Simulations were conducted using the Vienna ab initio simulation package[45]. The wave function was described by a plane-wave basis set with projected augmented wave method[46,47]. The exchange-correlation functional was simulated with the optB86b exchange function[48] and van der Waals density functional method[49,50], which was found to be accurate in reproducing layered structures[29, 51-53]. Energy cutoff for the plane wave basis was set to 400 eV in structural relaxations and increased to 500 eV in static calculations. Layered PtSe$_2$ and PtS$_2$ were modeled by a $1 \times 1$ supercell. 5L slab model was used for each type of materials, with a vacuum space of at least 20 Å in z direction. The k-mesh was sampled by a $3 \times 3 \times 1$ k-mesh, accuracy tested by a $5 \times 5 \times 1$ one. All atoms were relaxed until the residual force for each atom was less than 0.01 eV·Å$^{-1}$.

**ACKNOWLEDGMENTS**

We acknowledge the financial support from National Key Research & Development Program (No. 2016YFA0201902), the National Natural Science Foundation of China (No.

61604102, 51290273 and 11404372), ARC (DP140101501 and FT150100450), the Priority Academic Program Development of Jiangsu Higher Education Institutions (PAPD) and Collaborative Innovation Center of Suzhou Nano Science and Technology. Q. Bao acknowledges support from the Australian Research Council Centre of Excellence in Future Low-Energy Electronics Technologies (FLEET) (project number: CE170100039).

## ASSOCIATED CONTENT

**Supporting Information.** Details of the XRD spectra, XPS spectra, TEM images, UPS and KPFM measurements results, and supplementary photoelectrical results at 2200 nm, 532 nm and 1064 nm are provided. This material is available free of charge on the ACS Publication website at http://pubs.acs.org.

## AUTHOR INFORMATION

### Corresponding Author

Shenghuang Lin and Shaojuan Li

Tel: (+86)-512-65882337; Fax: (+86)-512-65880820; E-mail:

shenghuanglinchina@gmail.com; sjli@suda.edu.cn

### Author Contributions

Jian Yuan, Tian Sun and Zhixin Hu contributed equally to this work. The manuscript was written through contributions of all authors. All authors have given approval to the final version of the manuscript.

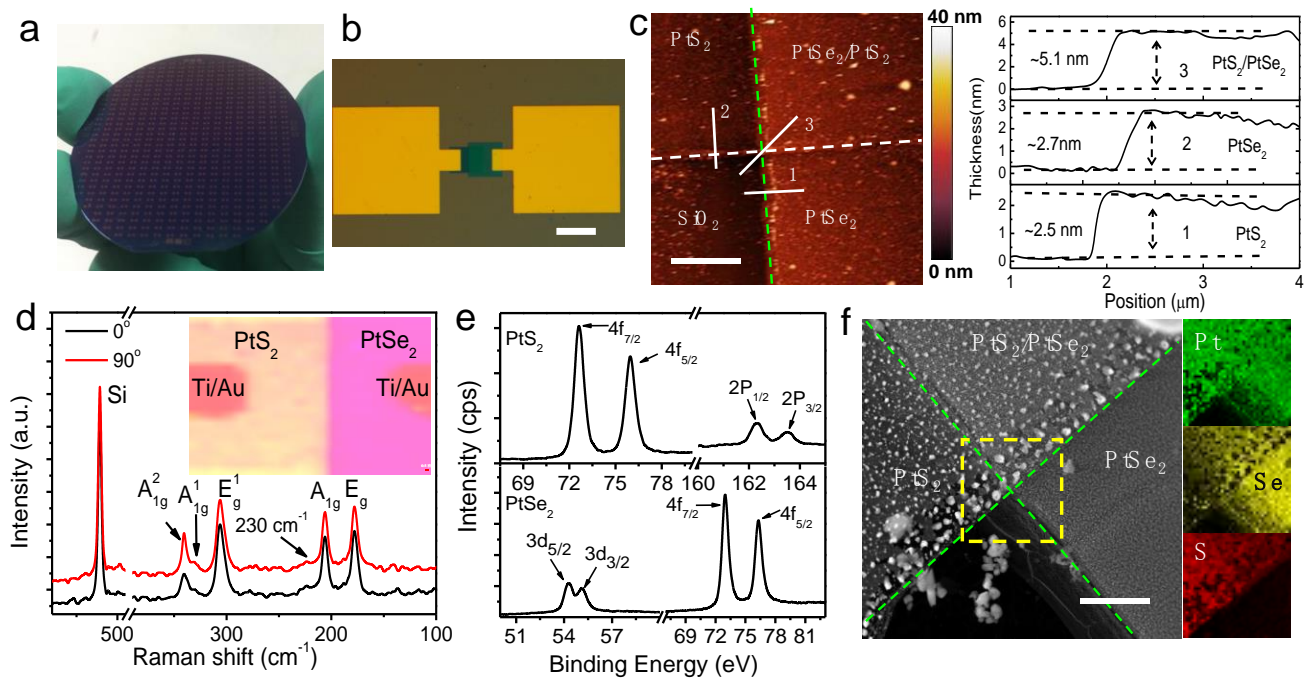

**Figure 1.** Material characterizations of the PtS$_2$/PtSe$_2$ heterojunctions. (a) The microscope image of PtS$_2$/PtSe$_2$ heterojunctions obtained on a 2″ SiO$_2$/Si wafer with 300 nm thick silicon dioxide. (b) The high magnification microscope image of PtS$_2$/PtSe$_2$ heterojunction device. Scale bar: 100 μm. (c) The AFM image and profile of PtS$_2$/PtSe$_2$ heterojunction. Scale bar: 5 μm. (d) The Raman spectra of PtS$_2$/PtSe$_2$ heterojunction. Inset: The corresponding Raman mapping image. (e). The high-resolution XPS spectra of PtS$_2$ and PtSe$_2$ film. (f) The bright-field STEM image of Pt, Se and S elements, respectively. Scale bar: 20 nm.

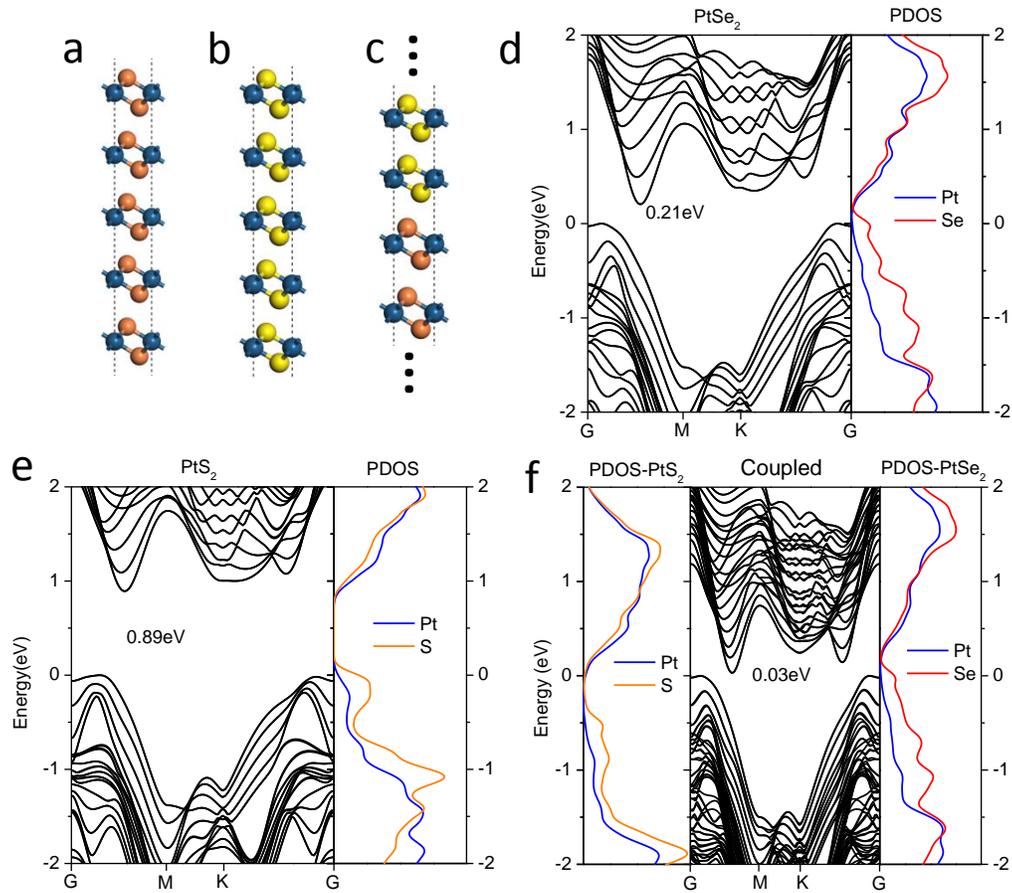

**Figure 2.** Simulations of the energy band of the PtS$_2$ and PtSe$_2$. (a-c) Atomic geometries and band structures of PtSe$_2$ and PtS$_2$. Blue, orange and yellow spheres in panel represent Pt, Se and S atoms respectively. 5L-PtSe$_2$ and 5L-PtS$_2$ are used in the coupled system in panel c. Some layers are not displayed in the figure in order to make the appearance clearer. The width of bandgap is inserted in band structure panels (d-f). Projected density of states (PDOS) for Pt and Se/S atoms are also shown adjacent to band structures.

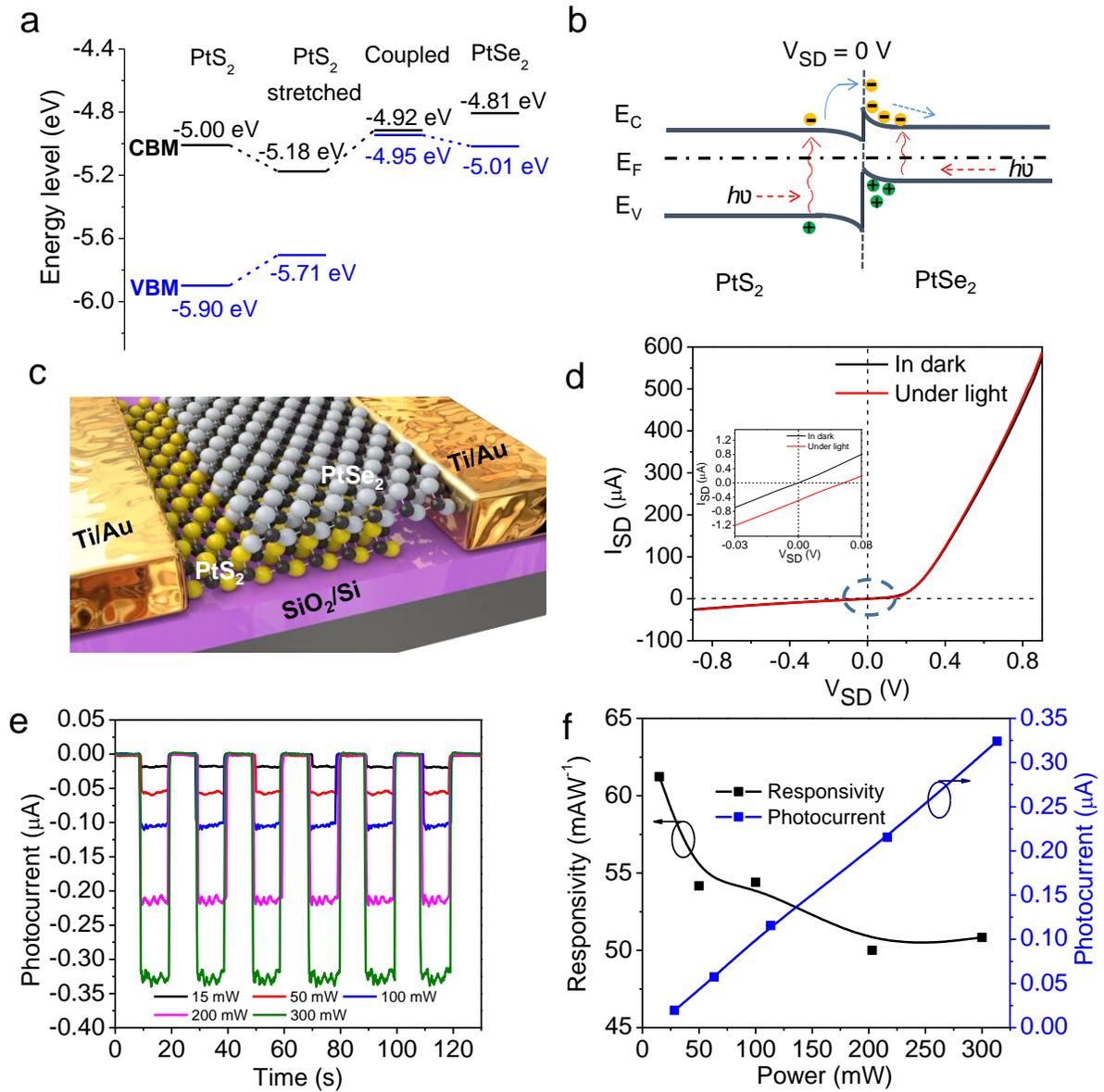

**Figure 3.** Energy diagrams and zero-bias photoresponse of the PtS$_2$/PtSe$_2$ heterojunctions. (a) Absolute energies of VBM (black) and CBM (blue) of PtS$_2$, PtSe$_2$ and the coupled structure. Dashed lines connecting those steps indicate the energy shift when the material is reshaped or coupled with other material. The column 'PtS$_2$ stretched' means lattice parameter of PtS$_2$ is stretched to adjust the value of PtSe$_2$. (b) Photoexcited carriers transfer process in the PtS$_2$/PtSe$_2$ heterojunctions under light illumination. E$_C$: conduction band; E$_V$: valence band. E$_F$: Fermi level. (c) Schematic illustration of a PtS$_2$/PtSe$_2$ heterojunction photodetector. (d) Representative *I-V* curves of the junction device at 1064 nm. Inset: Magnification image of *I-V* curves near zero bias voltage. (e) Time-dependent photocurrent

response exited under 1064 nm light illumination with different powers. (f) Dependence of photocurrent and responsivity on light power at 1064 nm. $V_{SD} = 0$ V, $V_G = 0$ V.

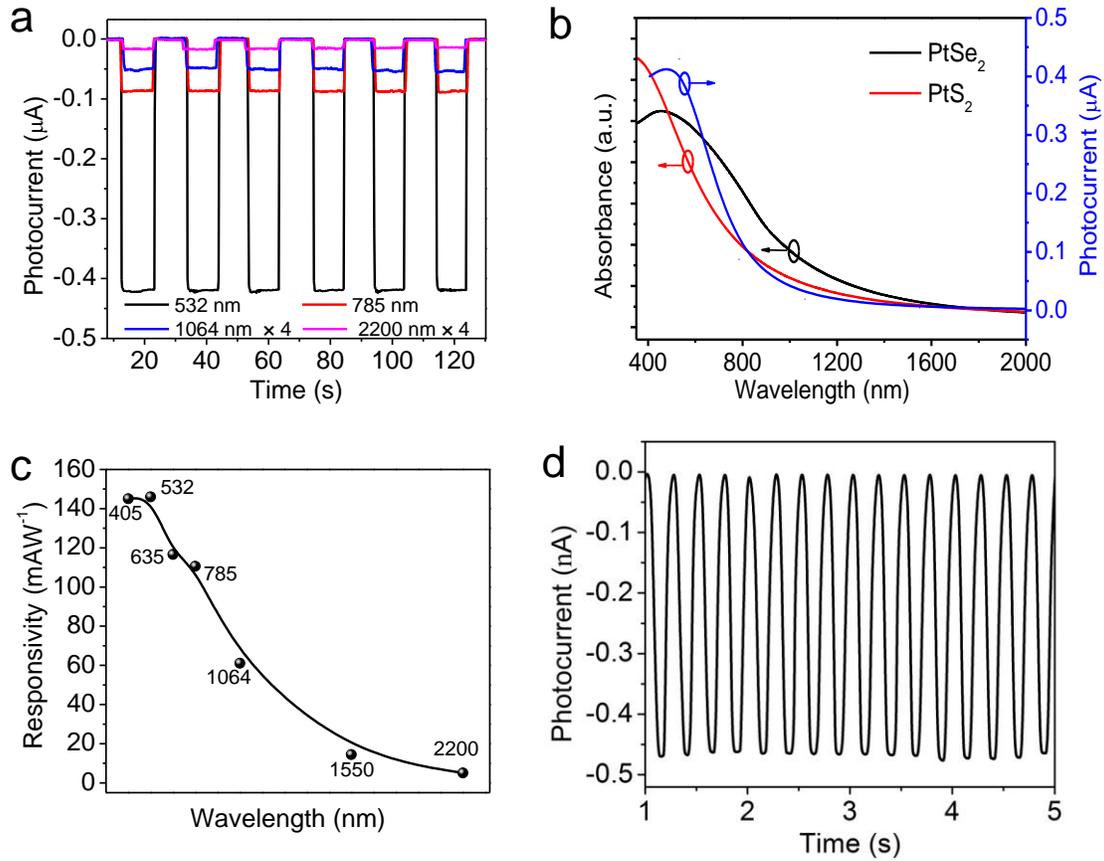

**Figure 4.** Broadband photoresponse of the PtS$_2$/PtSe$_2$ heterojunction devices. (a) The temperal photoresponse of the PtS$_2$/PtSe$_2$ heterojunctions with various wavelengths under the same light power (15 mW). $V_{SD} = 0$ V, $V_G = 0$ V. (b) The absorption spectrum of PtSe$_2$ and PtS$_2$ films and the corresponding wavelength-dependent photocurrent of the PtS$_2$/PtSe$_2$ heterojunction devices. (c) The wavelength-dependent responsivity under the same light power (15 mW). (d) Photoresponse of the PtS$_2$/PtSe$_2$ heterojunction devices to pulsed infrared light irradiation (1064 nm laser) with a frequency of 8 Hz. $V_{SD} = 0$ V, $V_G = 0$ V.

**Table 1** Comparison of the performance of PtS$_2$/PtSe$_2$ heterojunction devices with previously reported TMDs/TMDs and TMDs/BP heterojunctions.

| Materials | Measurement Conditions | Responsivity (mA/W) | EQE | Response time (ms) | Range of response wavelength (nm) | Reference |
|---|---|---|---|---|---|---|
| Multilayer MoS$_2$/WSe$_2$ | Vsd = 0 V 532 nm | ~120 | 34% | * | 500-800 | *Ref.4* |
| BP/MoS$_2$ | Vsd = 0 V 633 nm | * | 0.3% | * | * | *Ref.14* |
| Double gated MoSe$_2$ PN junction | Vsd = -1 V 532 nm | 0.7 | 0.1% | 10 ms | 400-800 | *Ref.32* |
| MoTe$_2$/MoS$_2$ | Vsd = 0 V 405 nm 800 nm | 322 38 | 88% 6% | 25 ms | 400-800 | *Ref.17* |
| PtS$_2$/PtSe$_2$ | Vsd = 0 V 1064 nm 1500 nm 2200 nm | 10.2 2.4 0.6 | 1.2 % 0.2 % 0.05% | 66 ms | 405-2200 | ***This work*** |